\documentclass[]{iopart}
\usepackage[T1]{fontenc}
\usepackage[latin1]{inputenc}
\usepackage{graphicx}

\begin{document}

\title[$\Lambda(1520)$ and $\Sigma(1385)$ production in
 Au+Au and p+p at $\sqrt{s_{NN}}=200$~GeV]
{$\Lambda (1520)$ and $\Sigma (1385)$ resonance production in Au+Au
and p+p collisions at RHIC ($\sqrt{s_{NN}}=200$~GeV)}

\author{Ludovic Gaudichet for the STAR collaboration %
\footnote{See \cite{Helen} for the full collaboration list.%
}}

\ead{Ludovic.Gaudichet@subatech.in2p3.fr}

\address{SUBATECH, École des Mines, 4 rue Alfred Kastler, 44307 Nantes, France}

\begin{abstract}
The production of $\Lambda (1520)$ in p+p and Au+Au collisions provided
by the RHIC at $\sqrt{s_{NN}}=200$ GeV are investigated using the
STAR detector. A preliminary $\Sigma (1385)$ signal is also shown.
Models predict that a late rescattering phase in ultra-relativistic
heavy ion collisions should lower the measured yields of these resonances.
This reduction is presently confirmed through the $\Lambda (1520)/\Lambda $
ratio which has been calculated for p+p collisions and compared with
Au+Au collisions.
\end{abstract}

\pacno{25.75.Dw}

\section*{Introduction}

Ultra-relativistic heavy ion collisions allow the study of created
hot and dense matter. The produced system is thought to evolve into
different stages before it breaks up. In particular an early partonic
stage is followed by the hadronization of the system. An important
rescattering phase may appear at the end of this scenario. It is of
the utmost importance to understand these final steps of heavy ion
collisions. Firstly because the study of this cooling medium will
constrain our models of heavy ion collisions and secondly because
most of our observations take their final shape during these phases.
Therefore the existence of specific observables which are able to
probe especially the hadronization and an eventual rescattering phase
are of great interest. Resonances produced during the hadronization
are such observables. Their short lifetime means that many of them
decay inside the hot and dense medium before the system breaks up.
Two types of effects are thought to appear. The reconstruction of
resonance signal through the decay tracks reveals the properties of
the resonance inside this medium. Mass modification or width broadening
have been expected \cite{Shuryak} and are reported by the STAR experiment
\cite{Patricia}. The second effect probes the existence of particle
rescattering in the hadronic gas. Since they are produced inside the
medium, decay tracks may undergo rescattering with other particles
and loose the momentum information on the decay. The yield of resonances
could therefore be reduced. The momentum dependence of the rescattering
may also modify the observed momentum distribution of resonances.
In order to investigate this medium effect, the production of $\Lambda (1520)$
and $\Sigma (1385)$ in p+p and Au+Au collisions are presented at
$\sqrt{s_{NN}}$ = 200 GeV, as measured by the STAR experiment. Models
investigating the rescattering phase expect indeed both resonances
to show a significant suppression in heavy ion collisions.

\section{Models}

\subsection{Strange hadron resonances within a statistical model and rescattering
effect}

\begin{figure}[tbh]
\begin{center}\includegraphics[  scale=0.31]{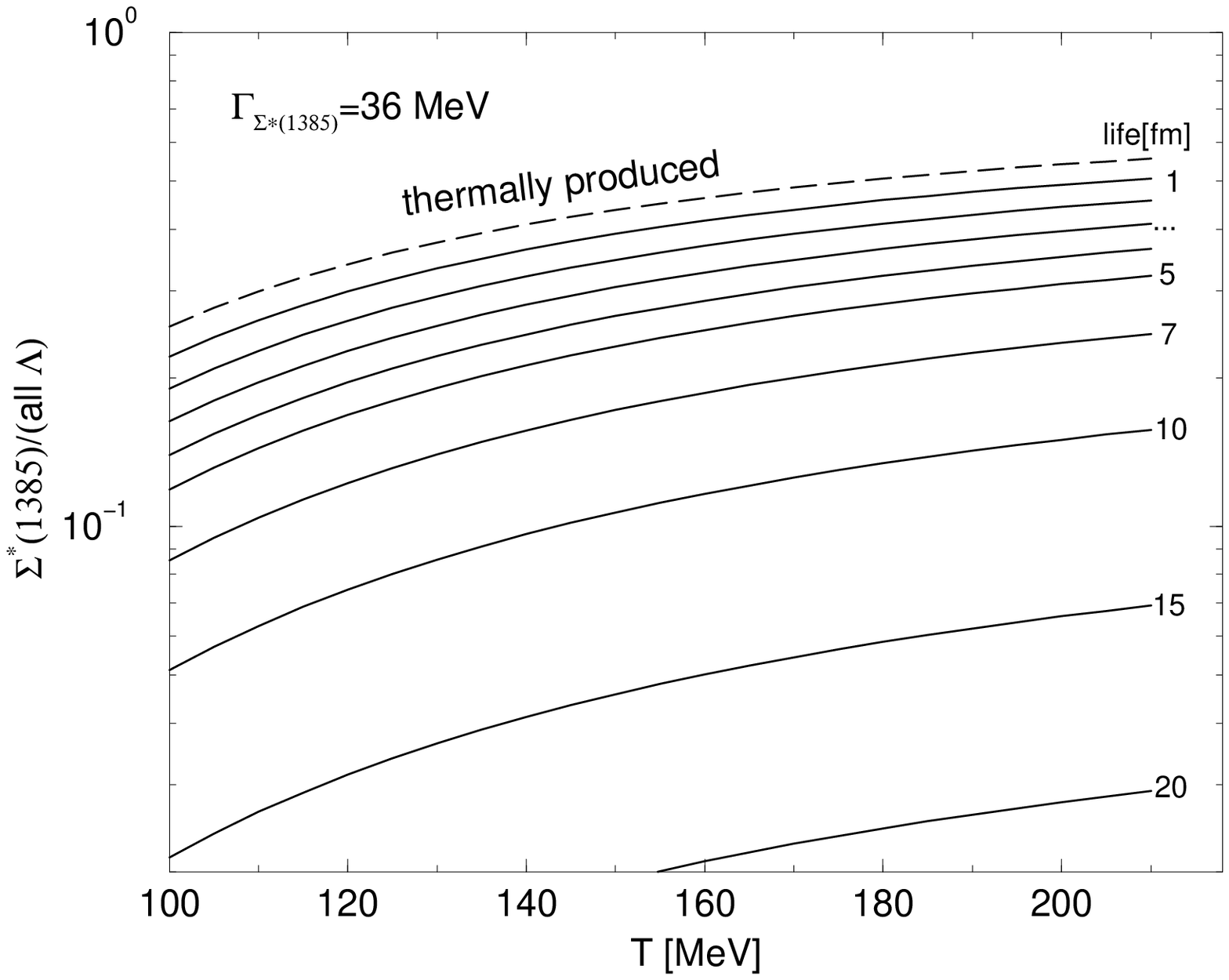}\includegraphics[  scale=0.3]{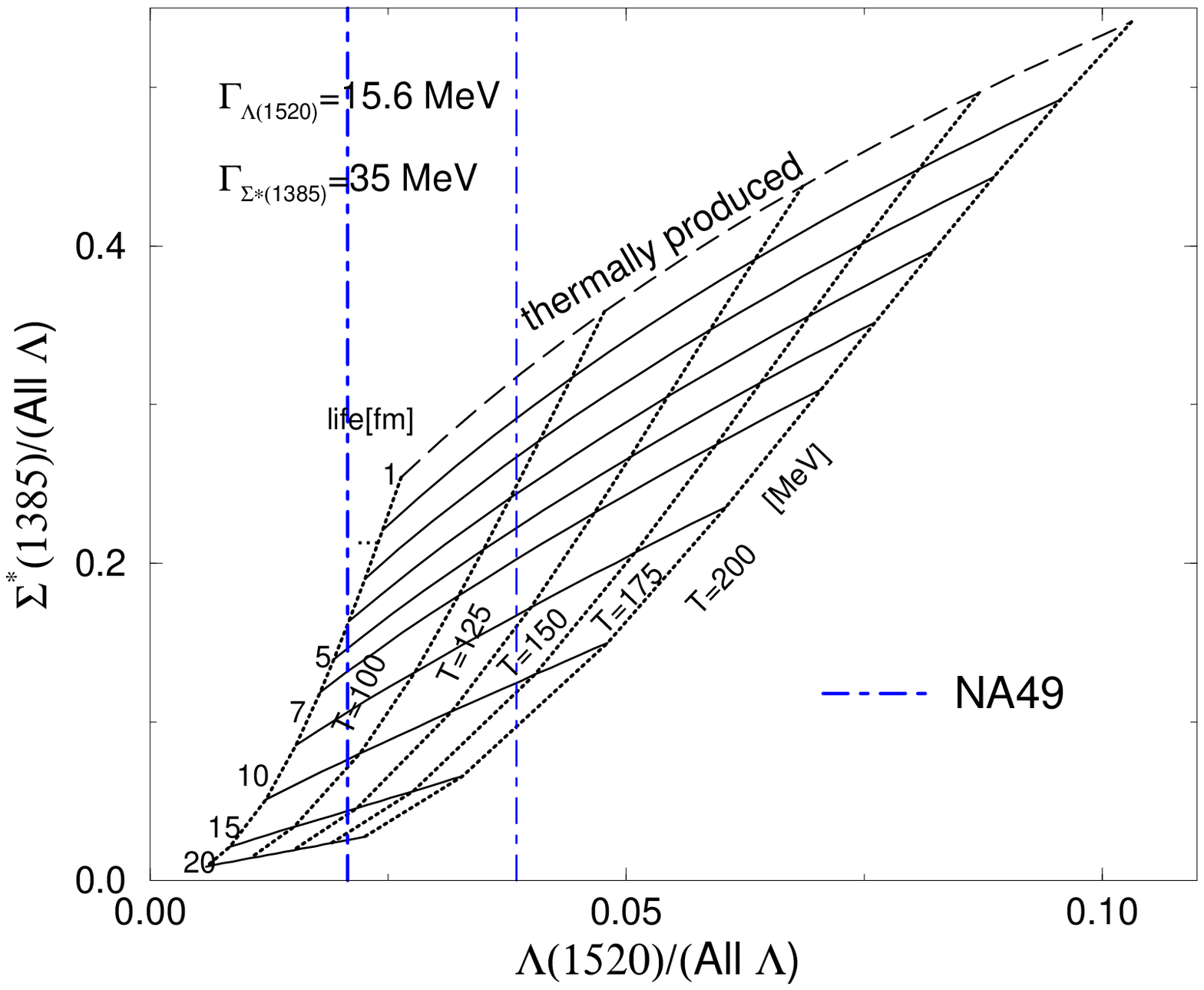}\end{center}

\caption{\label{cap:plotsRafelski}Left (a) : $\Sigma (1385)$/(All $\Lambda $)
ratio as a function of the temperature at chemical freeze-out and
the lifetime of the rescattering phase $\Delta t$. Right (b): $\Sigma (1385)/(all\, \Lambda )$
versus $\Lambda (1520)/(all\, \Lambda )$ diagram for different T
and $\Delta t$ ; (figures from \cite{Christina_Giorgio_Rafelski})}
\end{figure}
Strange hadron resonances could determine the time scale governing
hadron production and the duration of the decoupling phase. Ratios
of yields of resonances over the ground state particles with the same
valence quarks are estimated by a thermal model where the temperature
is the only parameter. However the rescattering of decay products
reduce the observed yields of resonances. This process can be take
into account by a microscopic model which calculates the suppression
of the resonance signal during the hadronic phase. Thus coupling the
thermal production and the rescattering characterization allows one
to calculate these ratios as a function of the time between chemical
and thermal freeze-out ($\Delta t$) versus the temperature at chemical
freeze-out T \cite{Torrieri_rafelski}. Figure \ref{cap:plotsRafelski}
(a) shows the $\Sigma (1385)$/(All $\Lambda $) ratio as a function
of the temperature and $\Delta t$. figure \ref{cap:plotsRafelski}
(b) demonstrates that using at least two resonances with different
lifetimes gives a unique value for both T and $\Delta t$. However
the model is not complete since it does not take into account the
resonance regeneration by re-interaction of decay products in the
hadronic gas.

\subsection{Probing chemical and thermal freeze-outs via UrQMD}

The microscopic transport approach of the Quantum Molecular Dynamics
model (UrQMD) \cite{UrQMD1} can directly address the question of
the production and observability of resonances \cite{Bleicher_Aichelin}.
Dynamics are described in terms of inelastic and (pseudo) elastic
collision rates. After an early non equilibrium stage, the system
is dominated by inelastic collisions and chemistry changing processes.
The elastic and pseudo-elastic collisions become dominant in the next
stage where they mostly change the momenta of the hadrons. The model
shows therefore a separation between a chemical freeze-out and thermal
freeze-out, i.e. a period of time when resonance signal can be destroyed.
The lowering of the resonances observability can directly be calculated
by dividing the number of resonances for which decay products have
survived the rescattering phase by the number of all produced resonances.
It appears that 30\% of $\Lambda (1520)$ produced in UrQMD at $\sqrt{s_{NN}}=200$
GeV should not be observable due to the rescattering of at least one
of the decay products. Part of this suppression comes from a double
counting process. A resonance which decays and then is regenerated
by a reaction involving one of its decay products is counted twice
in the number of generated resonances but once in the number of suppressed
resonances.

\section{Data analysis}

Table \ref{cap:Reconstruct-decay-channel} shows the decay channels
which are used in the reconstruction of $\Lambda (1520)$ and $\Sigma (1385)$.
The large Time Projecting Chamber (TPC) of STAR has been used to detect
the charged tracks at mid-rapidity. Due to the short lifetime of these
resonances, the secondary vertex of their decay can not be separated
from the primary vertex of the collision. Signal reconstruction is
therefore obtained by associating all selected tracks of the daughter
species which seem to come from the primary interaction. Reconstruction
of $\Sigma (1385)$ is done via combinations of three tracks, the
pion of the resonance decay and the two decay tracks of the reaction
$\Lambda \rightarrow p\pi ^{-}$(64\%). The large combinatorial background
is reproduced by event mixing \cite{event-mixing}. Flow in heavy
ion collisions or jet processes in p+p collisions introduce an azimuthal
asymmetry to the analyzed events. This asymmetry leads to a characteristic
shape on mixed event distributions. This behavior has been corrected
for all invariant mass distributions of $\Lambda (1520)$ \cite{maStarNote}.

Seven million minimum bias p+p events have been used to produce the
$\Sigma (1385)$ signal. The $\Lambda (1520)$ analyses use 10 million
p+p events, 1.7 million minimum bias Au+Au events and 1.7 million
central trigger Au+Au events. Au+Au minimum bias events are divided
in four centrality categories from peripheral to most central events~:
80-60\%, 60-40\%, 40-10\%, and 10\% or less of the hadronic cross
section.

\begin{table}
\begin{center}\begin{eqnarray*}
\Lambda (1520) & \rightarrow  & pK^{-}\, \, \, (22.5\%)\\
\Sigma (1385)^{-} & \rightarrow  & \Lambda \pi ^{-}\, \, \, (88\%)\, \, \, \, (\textrm{decay mode shared with the }\Xi ^{-})\\
\Sigma (1385)^{+} & \rightarrow  & \Lambda \pi ^{+}\, \, \, (88\%)
\end{eqnarray*}
\end{center}

\caption{\label{cap:Reconstruct-decay-channel}Observable decay channels in
STAR for $\Lambda (1520)$ and $\Sigma (1385)$ resonances.}
\end{table}

\section{Results}

\subsection{$\Lambda (1520)$ and $\Sigma (1385)$ in p+p collisions}

\begin{figure}
\begin{flushright}\includegraphics[  scale=0.23]{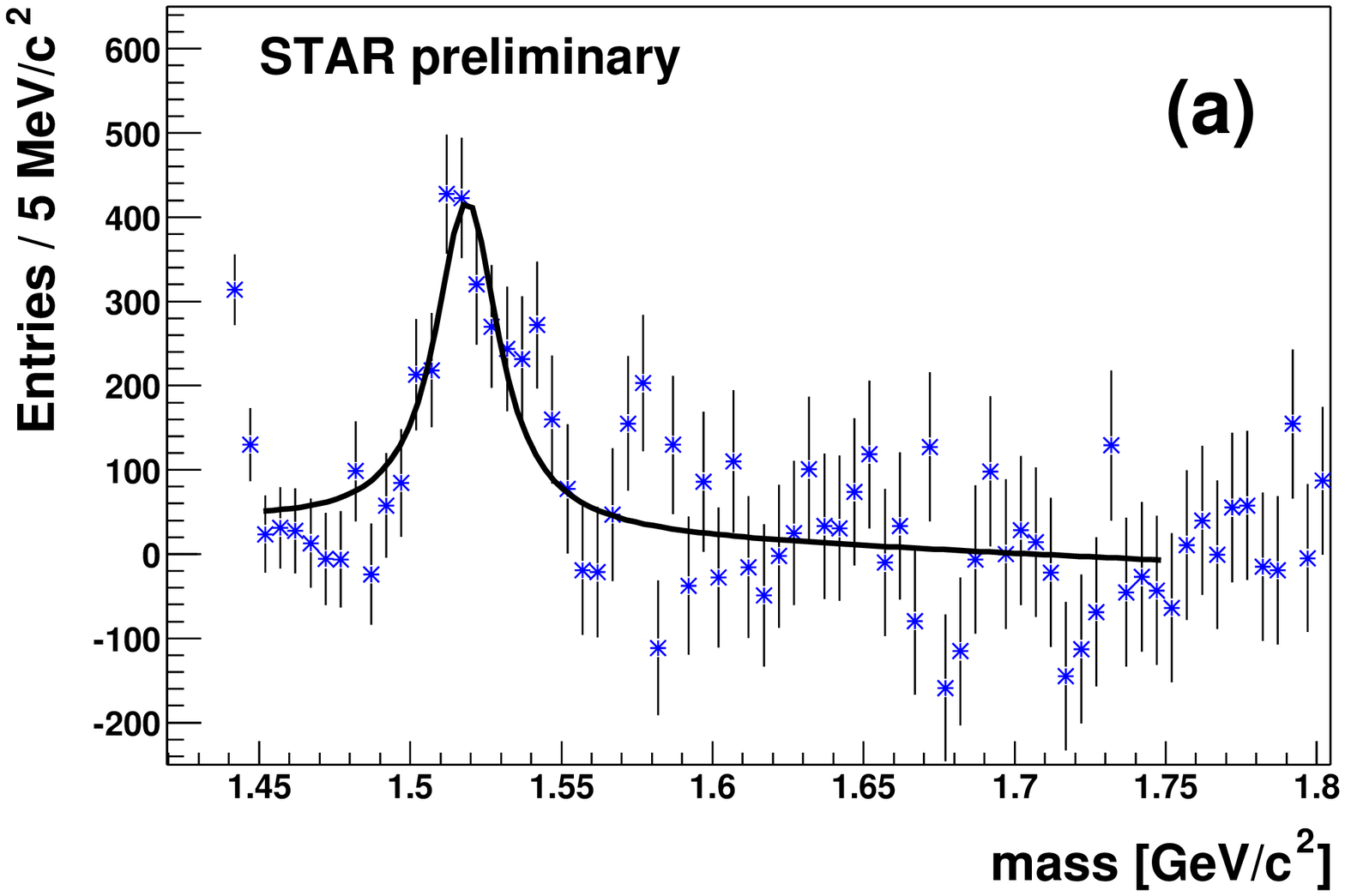}\includegraphics[  scale=0.23]{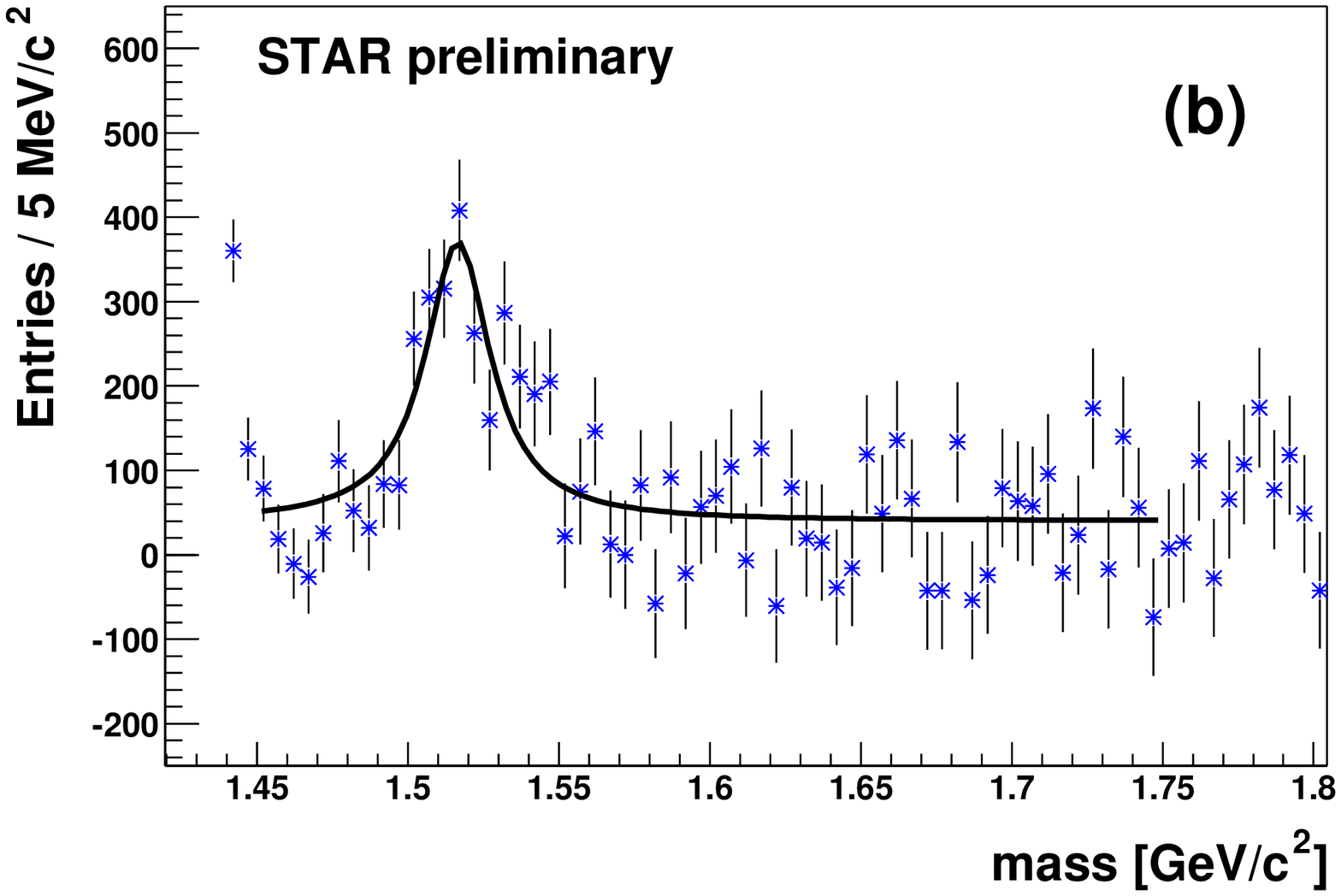}\includegraphics[  scale=0.23]{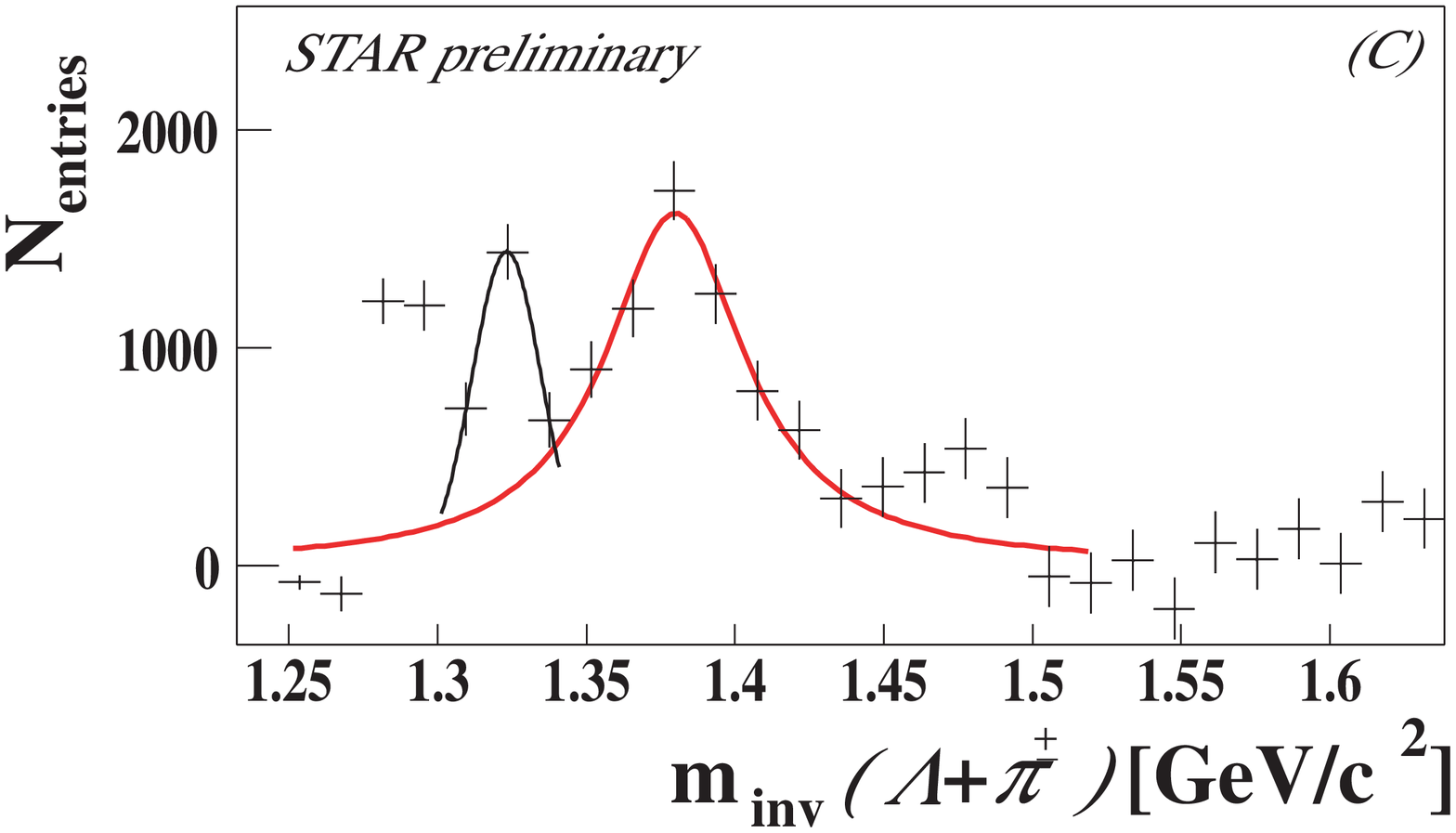}\end{flushright}

\caption{\label{cap:ppMinvs}Invariant mass distribution of (a) $\Lambda (1520)$,
(b) $\overline{\Lambda }(1520)$ and (c) $\Sigma (1385)+\overline{\Sigma }(1385)$
in p+p collisions.}
\end{figure}
\begin{figure}
\begin{center}\includegraphics[  scale=0.28]{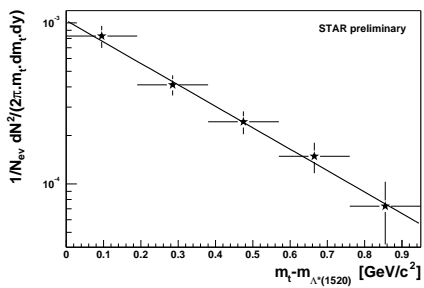}\end{center}

\caption{\label{cap:MtLAL1520inPP}Transverse mass distribution of $\left(\Lambda (1520)+\overline{\Lambda }(1520)\right)/2$
for the p+p collisions. Errors shown are statistical only.}
\end{figure}

Figure \ref{cap:ppMinvs} shows separately the invariant mass distributions
of $\Lambda (1520)$ and $\overline{\Lambda }(1520)$ in p+p collisions.
The mass and width are respectively $m=1518\pm 2$~$MeV/c^{2}$ and
$\Gamma =25\pm 5$~$MeV/c^{2}$ from a Breit-Wigner plus linear background
fit. The mass agrees with the Particle Data Group value of $1519.5\pm 1$~$MeV/c^{2}$
\cite{PDG}. The experimental width includes the $\Lambda (1520)$
natural width of 15.6~$MeV/c^{2}$ and the momentum resolution. Within
the errors this broadening of the width agrees to the Monte Carlo
simulations of the $\Lambda (1520)$ reconstruction. The raw ratio
$\overline{\Lambda }(1520)/\Lambda (1520)$ is equal to $0.90\pm 0.11(stat)$.
Both $\Lambda (1520)$ and $\overline{\Lambda }(1520)$ signals have
been added to decrease the statistical errors in the transverse mass
distribution ($m_{T}=\sqrt{P_{T}²+m_{0}²}$) of figure \ref{cap:MtLAL1520inPP}.
In order to correct raw yields, one Monte Carlo $\Lambda (1520)$
has been embedded per real p+p event. The analysis procedure has then
been applied on the simulated resonances to calculate the efficiency
and acceptance of the detector. The corrected transverse mass distribution
of $\left(\Lambda (1520)+\overline{\Lambda }(1520)\right)/2$ has
been fit to an exponential function in order to extrapolate the $dN/dy=0.0039\pm 0.0003(stat)$
and the slope parameter $T=326\pm 42(stat)$ MeV. Systematic errors
are determined by varying cuts, signal extraction and fit procedures.
Systematic error on yield is 15\% while the systematic error on the
slope parameter is about 30\%.

Preliminary invariant mass distribution of $\Sigma (1385)$+$\overline{\Sigma }(1385)$
is shown in figure~\ref{cap:ppMinvs} (c). The left peak is due to
the $\Xi $ and $\overline{\Xi }$ hyperons at the measured mass of
$1321\pm 1$~$MeV/c^{2}$. The right peak is fit to a Breit-Wigner
plus a linear background function. The mass and width are $m=1381\pm 2$~$MeV/c^{2}$
and $\Gamma =58\pm 7$~$MeV/c^{2}$. The integrated raw yield of
$\Sigma (1385)$+$\overline{\Sigma }(1385)$, extracted from the fit,
is about $7500\pm 100$ particles out of the 7 million minimum bias
p+p events, with a 20\% systematic error.

\subsection{$\Lambda (1520)$ in Au+Au collisions}

\begin{figure}
\begin{flushright}\includegraphics[  scale=0.23]{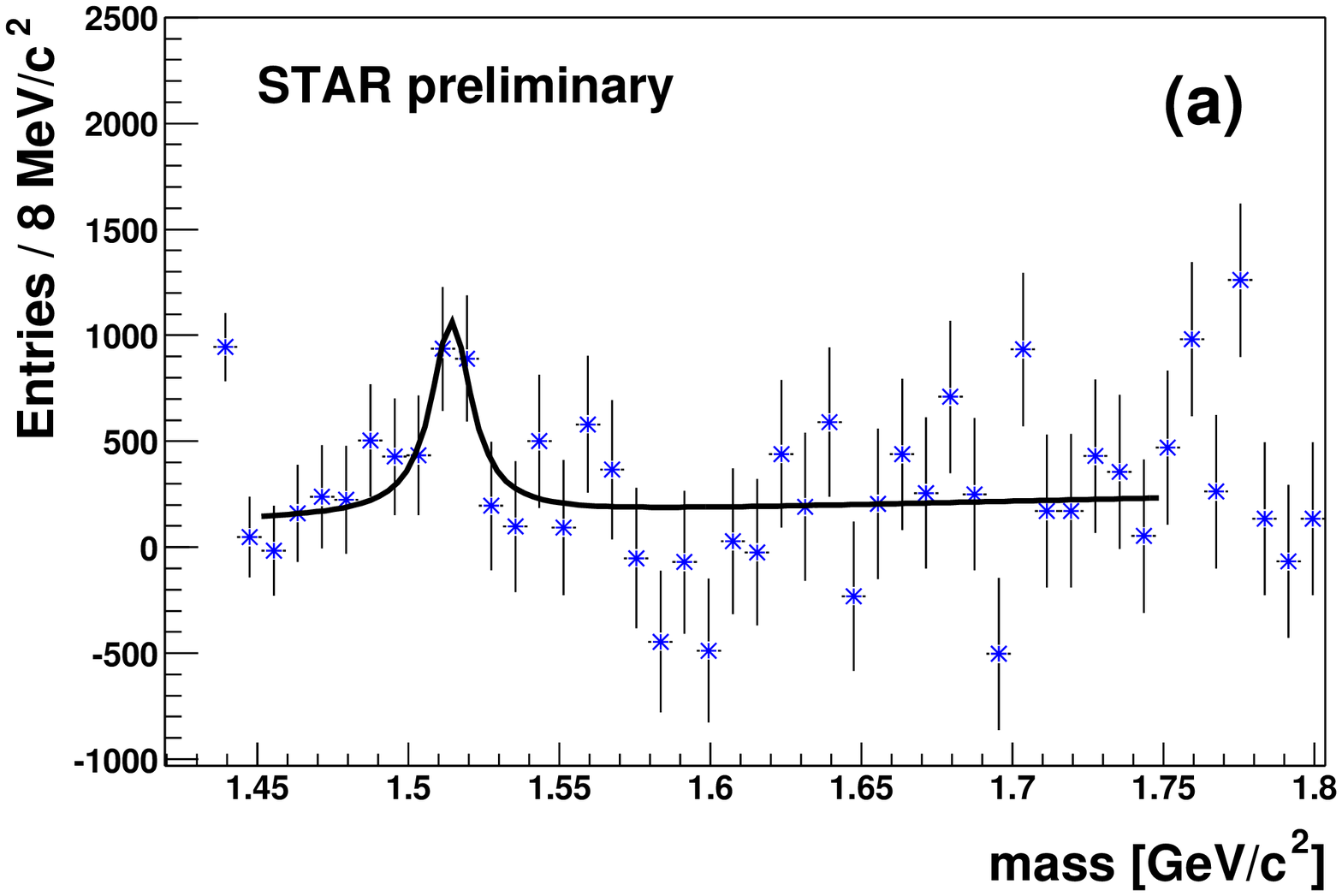}\includegraphics[  scale=0.23]{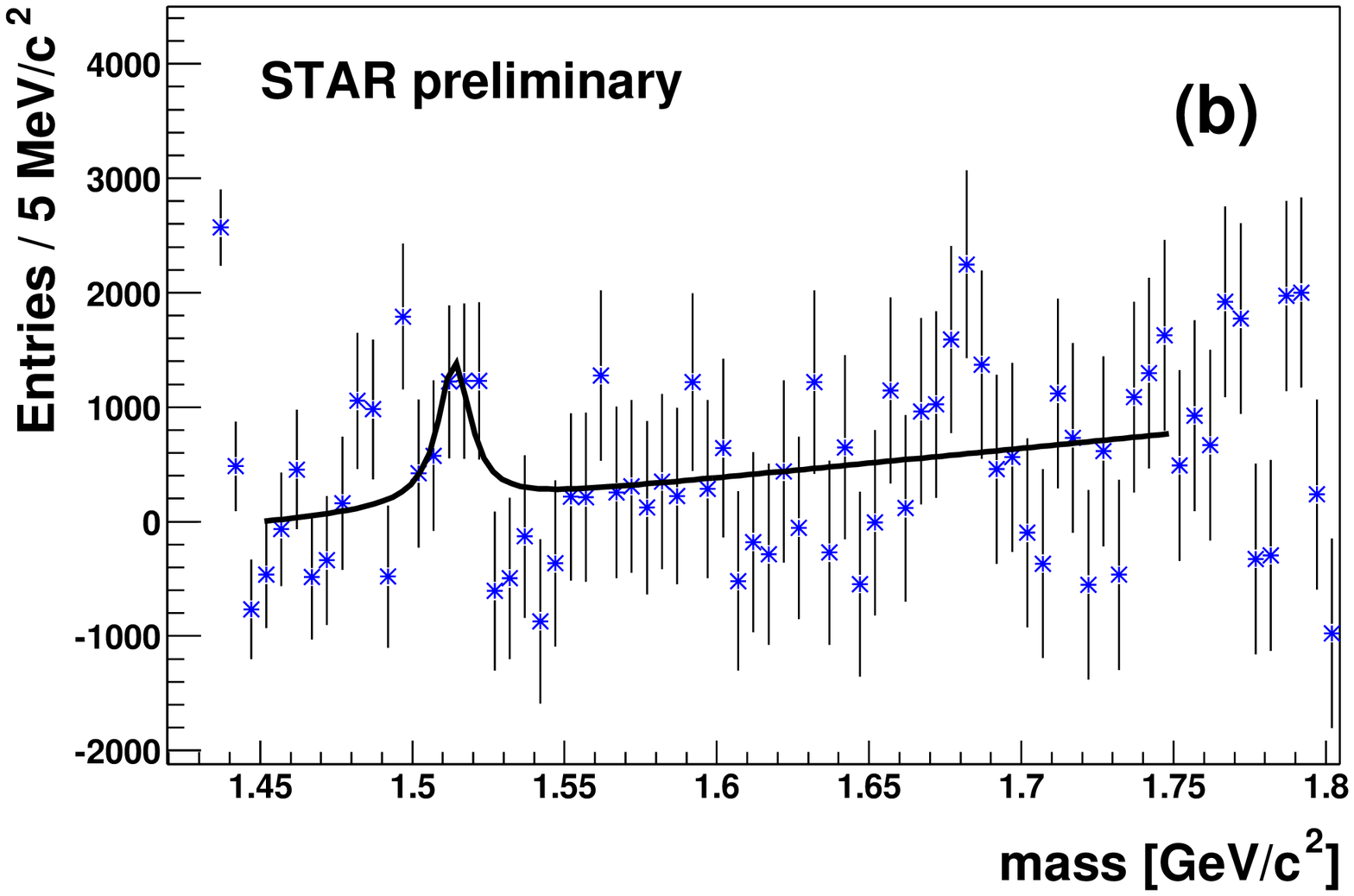}\includegraphics[  scale=0.18]{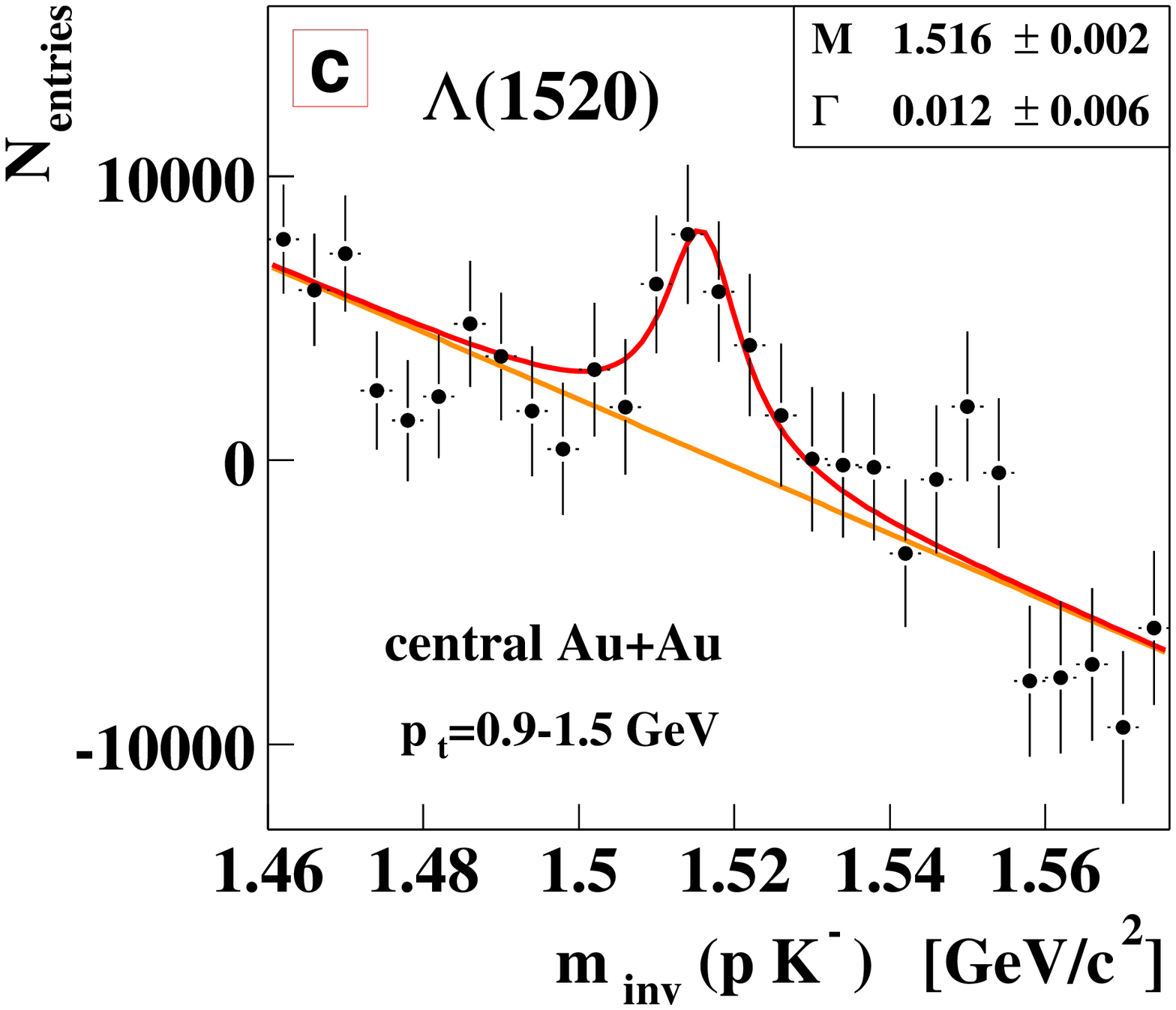}\end{flushright}

\caption{\label{cap:LAL1520inAuAuPeriph}$\Lambda (1520)+\overline{\Lambda }(1520)$
invariant mass for (a) the 80-60\% and (b) the 60-40\% centrality
categories of minimum bias Au+Au events. $\Lambda (1520)$ invariant
mass distribution for the STAR central events (c).}
\end{figure}
The gold minimum bias data set shows a $\Lambda (1520)$ signal for
the two most peripheral categories. Figure \ref{cap:LAL1520inAuAuPeriph}
(a) and (b) shows the invariant mass distributions. $\Lambda (1520)$
and $\overline{\Lambda }(1520)$ signals are added to be more statistically
significant. The 80-60\% distribution has a three sigma signal while
the 60-40\% distribution shows a two sigma signal. Transverse mass
distributions can not be performed due to the small numbers of detected
resonances. Integrated yields have therefore been calculated assuming
exponential $m_{T}$ distributions. Slope parameters, extracted from
$\Lambda $ analyses, are $T=260$ MeV for the 80-60\% signal and
$T=290$ MeV for the 60-40\% signal. $\Lambda (1520)$ and $\overline{\Lambda }(1520)$
mid-rapidity production is $0.086\pm 0.039$ per unit of rapidity
per event for the 80-60\% category. The yield is $0.12\pm 0.1$ for
the 60-40\% category. Quoted errors include 30\% systematic errors.
Figure \ref{cap:LAL1520inAuAuPeriph} (c) shows central collision
measurement of the yield $\left\langle \Lambda (1520)\right\rangle _{|y|<0.5}=0.58\pm 0.021(stat)\pm 30\%(sys)$
\cite{ChristinaAtBreckenridge}.

\subsection{$\Lambda (1520)/\Lambda $ ratios}

\begin{figure}
\begin{center}\includegraphics[  scale=0.28]{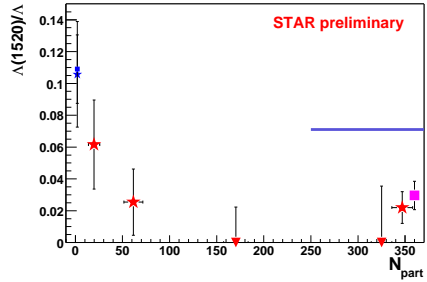}\end{center}

\caption{\label{cap:L1520overLambdaRatios}ratio $\Lambda (1520)/\Lambda $
as a function of the number of participants and thermal prediction
(line) from \cite{Dan}.}
\end{figure}

In order to estimate the predicted reduction of $\Lambda (1520)$
signal in heavy ion collisions, the ratio $\Lambda (1520)/\Lambda $
has been calculated for Au+Au collisions at different centralities
and is compared to the same ratio for p+p collisions. Figure \ref{cap:L1520overLambdaRatios}
includes all $\Lambda (1520)/\Lambda $ ratios as a function of the
number of participants. Square points show NA49 ratios in p+p and
Pb+Pb collisions at $\sqrt{s}$ = 17 GeV \cite{FrieseQM2002}. Other
points are STAR measurements. $\Lambda (1520)/\Lambda $ ratios from
the two most peripheral categories of Au+Au collisions are represented
at $<N_{part}>\sim $20 and $<N_{part}>\sim $62. Arrows show upper
limits at 95\% of confidence level for more central categories of
minimum bias events where no $\Lambda (1520)$ signal is visible.
A thermal prediction from \cite{Dan} is also represented by the horizontal
line. The ratio shows a significant decrease from p+p collisions to
peripheral Au+Au collisions. Furthermore the production of $\Lambda (1520)$
in heavy ion collisions is lower than the yield predicted by a thermal
model.

\section*{Summary}

We report preliminary results on the production of $\Lambda (1520)$
resonances in p+p collisions and in centrality categories of Au+Au
heavy ion collisions at RHIC at 200 GeV. $\Sigma (1385)$ signal is
also shown in p+p collisions. The $\Lambda (1520)/\Lambda $ ratio
have been determined for the different systems. Results agree with
NA49 data and show a decrease of the $\Lambda (1520)$ yield relative
to the $\Lambda $ yield. Already lower than the p+p value in peripheral
collisions, the $\Lambda (1520)/\Lambda $ ratio appears to be significantly
lower than the thermal expectation. These results are in qualitative
agreement with model predictions of a lowering of resonance production
in ultra-relativistic heavy ion collisions, due to the rescattering
of decay products in the medium.

\ack{I would like to thank Christina Markert and Sevil Salur for their
contributions to this work and Patricia Fachini for helpful discussions.
STAR acknowledgements can also be found in \cite{Helen}.}

\end{document}